\documentclass[aps,onecolumn,showpacs,floatfix, 12pt]{revtex4}
\usepackage[T1]{fontenc}
\usepackage{graphicx}
\usepackage[usenames]{color}

\begin{document}
\title{
A study of atom localization in an optical lattice by analysis of the
scattered light
}
\author{C. I. Westbrook, C. Jurczak$^1$, G. Birkl$^2$, B. Desruelle, W. D. Phillips$%
^3 $, A. Aspect}
\affiliation{Institut d'Optique, UA14 du CNRS, B.P.147 91403 Orsay CEDEX}

\begin{abstract}
We present an experimental study of a four beam optical lattice using the
light scattered by the atoms in the lattice. We use both intensity
correlations and observations of the transient behavior of the scattering
when the lattice is suddenly switched on. We compare results for 3 different
configurations of the optical lattice. We create situations in which the
Lamb-Dicke effect is negligible and show that, in contrast to what has been
stated in some of the literature, the damping rate of the 'coherent' atomic
oscillations can be much smaller than the inelastic photon scattering rate.
\end{abstract}

\maketitle

\section{Introduction}

In the past five years, optical lattices have attracted a great deal of
experimental and theoretical interest \cite{jessen2}. Beginning with studies
of the quantized motion of atoms in the sub-micron sized potential wells 
\cite{verkerk1,jessen1}, many workers have succeeded in observing a large
variety of effects and are now beginning to investigate the transport
properties, both classical and quantum in these systems \cite
{courtois2,jurczak2,anderson,marksteiner2}.

There are several methods of observation of optical lattices, and among them
two groups, including ours, recently demonstrated that intensity
correlation, a well established technique in other fields\cite{Chu}, can be
used to gain information\cite{jurczak1,bali}. Although the bulk of our new
results has been related to the transport properties of atoms in these
lattices (i.e. the motion from well to well) \cite{jurczak2}, our
experiment, like many others, also gives information about the motion of
atoms inside the wells. In this paper we will detail our results on this
motion and use our data to confirm recent ideas \cite{grynberg,comment}
about the damping mechanisms which affect the motion of atoms inside the
lattice potential wells.

A novel feature in our experiment is that we can study the behavior of atoms
in a tetrahedral optical lattice for a large range of beam angles. This
feature has already permitted us to study transport by density waves in
lattices as a function of lattice angle\cite{jurczak3}. Here we will discuss
a study of the intrawell dynamics, i.e. the oscillation of atoms inside a
single well as a function of lattice angle.

In addition to the intensity correlation measurements, we have also studied
the same dynamics by observing the atomic fluorescence as a function of time
after a homogeneous atomic sample is suddenly subjected to an optical
lattice. Although not as powerful a method as Bragg scattering for studying
the localization of atoms in individual wells, we demonstrate in the
appendix that a fairly simple experiment is indeed sensitive to the degree
of localization in a single well.

\section{Theoretical Remarks}

In our experiments we used a 4-beam optical lattice configuration introduced
in Ref.~\cite{verkerk2}. The configuration of laser beams and polarizations
is shown in Fig. 1. We denote by $\theta $ the half angle between the
lattice beams in the $x-z$ and $y-z$ planes. The $x-z$ and $y-z$ angles were
always the same. The resulting electric field and optical potential are
discussed in detail in Ref. \cite{petsas}. Here we will simply state the
results of this work which are important for our study. For an atom with an F%
$_{g}$=3 ground state and an F$_{e}$=4 excited state, Montecarlo
wavefunction simulations\cite{marksteiner}\cite{raithel} have shown that
atoms in the lattice are rapidly pumped into the extreme magnetic sublevels $%
m_{F}=\pm 3.$ Therefore we will neglect the potentials not corresponding to
these two states. The three dimensional light shift potential for atoms in
the m$_{F}$ =$\pm $3 state is given by: 
\begin{equation}
U_{\pm }=-\frac{U_{0}}{4}\left( \frac{29}{28}\left( \cos ^{2}(k_{x}x)+\cos
^{2}(k_{y}y)\right) \mp \frac{27}{28}\left( 2\cos (k_{x}x)\cos (k_{y}y)\cos
(2k_{z}z)\right) \right)  \label{potential}
\end{equation}
Where $k_{x}=k_{y}=k\sin \theta ,$ $k_{z}=k\cos \theta $ , $k=2\pi /\lambda $
is the wavevector of the light and $U_{0}$ is the light shift at the bottom
of the potential well. The numerical factors come from the values of the
Clebsch Gordan coefficients. In the large detuning limit this light shift is
related to the saturation parameter $s_{0}=\frac{\Omega _{R}^{2}}{2}/\left(
\Delta ^{2}+\frac{\Gamma ^{2}}{4}\right) $ of a single laser beam by $%
U_{0}=4\hbar \Delta s_{0},$ with $\Omega _{R}$ being the Rabi frequency, $%
\Delta $ the detuning and $\Gamma $ the natural linewidth. From the
curvature of the bottom of these wells one can calculate an oscillation
frequency along each axis for an atom near the bottom of each well: 
\begin{eqnarray}
\Omega _{x,y} &=&\omega _{R}\sqrt{\frac{2U_{0}}{E_{R}}}\sin \theta
\label{oscfreq} \\
\Omega _{z} &=&2\omega _{R}\sqrt{\frac{28}{27}\frac{U_{0}}{E_{R}}}\cos \theta
\end{eqnarray}
Here $E_{R}=\frac{\hbar ^{2}k^{2}}{2m}$ denotes the recoil energy, and $%
\omega _{R}=E_{R}/\hbar .$
%
% Figure 1
%
\begin{figure}[b]
\begin{center}
\includegraphics[width=10 cm]{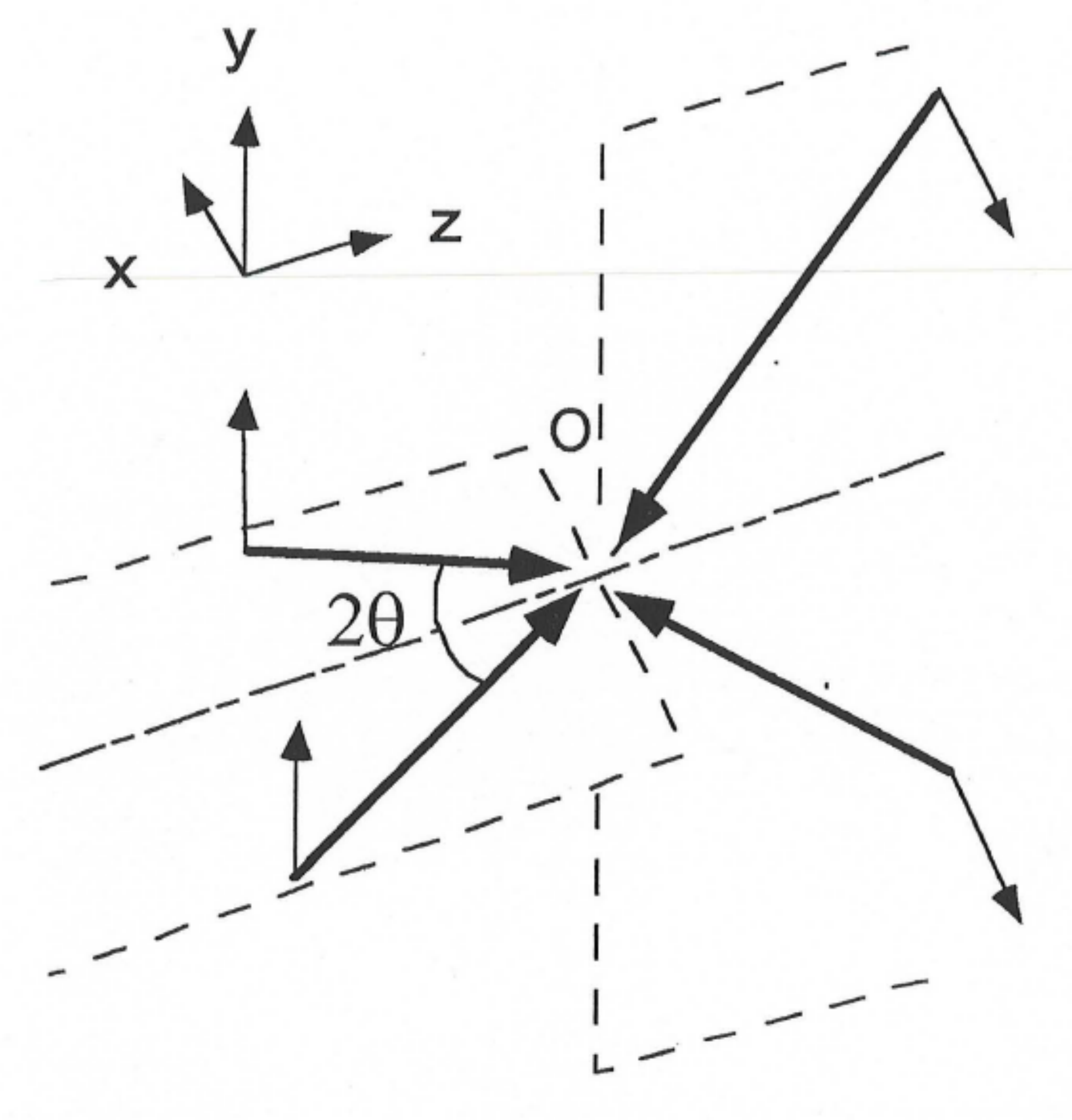}
\caption{Configuration of laser beams and polarizations used to create the
optical lattice used in this experiment. The heavy arrows show the beam
propagation directions and the fine arrows the polarizations.}
\label{figure-1}
\end{center}
\end{figure}

These oscillation frequencies have been observed using many different
methods. Our intensity correlation technique amounts to observing the beat
note between the light radiated by different atoms. Since the atoms are
oscillating at well defined frequencies, the power spectrum of the emitted
electric field contains a 'carrier' at the laser frequency and sidebands
separated by multiples of the oscillation frequency. Thus the
'self-beating', or intensity correlation spectrum contains beat notes at
these frequencies. Instead of self-beating one can also use a local
oscillator (heterodyning) to observe the sidebands 
\cite{jessen1,gatzke,westbrook}. 
Both of these methods can also be considered as
spontaneous Raman spectroscopy of the vibrational levels in the potential
wells. In pump-probe spectroscopy, on the other hand, \cite
{verkerk1,hemmerich2} one observes stimulated Raman transition between the
energy levels in the wells. Four-wave mixing signals also contain motional
sidebands \cite{hemmerich}. Recently various groups have observed
oscillations in a transient way, by non-adiabatically changing the potential
wells and observing the motion either through the redistribution of photons
in the lattice beams \cite{kozuma,phillipspc} or using Bragg scattering \cite
{raithel,raithel2,gorlitz}.

In the first discussions of the motion of atoms inside the potential wells
of an optical lattice, attention was drawn to the fact that the motional
sidebands, which were the signature of a well defined oscillation of the
atoms in the wells, appeared to be narrower than the simplest expectation 
\cite{verkerk1,jessen1}. Naively, one expects the linewidth of such a system
to be approximately equal to the total photon scattering rate (at the bottom
of a well this rate is given approximately by $\frac{U_0}{\hbar \Gamma }%
\Delta $). References \cite{verkerk1,jessen1}, however, pointed out that in
fact only the inelastic scattering rate (the rate of scattering events
involving a change of the quantum state of the atom) should determine that
linewidth. Because the experiments were in the Lamb-Dicke regime, i.e. the
amplitude of the oscillation was much smaller than the wavelength of the
emitted radiation, inelastic scattering was strongly suppressed in these
experiments \cite{wineland,courtois}, and this accounted for the observed
width.

This argument is roughly correct if there is no significant excitation of
the harmonic oscillator, i.e. most of the atoms are in the ground state, as
was indeed approximately true in the experiments of Refs. \cite
{verkerk1,jessen1}. Later, however, Ref. \cite{grynberg} pointed out that
when the amplitude of the atomic oscillation is large enough that one must
include the effects of many levels in the well, there is a transfer of
coherence mechanism which suppresses the linewidth even further.
Essentially, this mechanism involves the fact that although coherences
between adjacent levels in a well do decay at a rate determined by the
inelastic scattering rate, these coherences are also 'fed' by other, higher
lying coherences, provided that the higher lying coherences oscillate at
approximately the same frequency as the lower lying ones, i.e. that the well
is nearly harmonic.

Indeed, in the case of a perfectly harmonic well, Ref.~\cite{cct}
demonstrates that the coherence transfer mechanism works in such a way that
the linewidth of the oscillator is determined only by the rate at which
energy is extracted from the oscillator through its coupling to a thermal
reservoir, and not by the decay rates of the individual coherences. This
extraction rate is exactly the same as the damping, or cooling rate of the
classical oscillator.

The case of the perfectly harmonic oscillator was adapted to the case of
laser cooling in Ref.~\cite{comment} to explain the damping rates observed
in an experiment reported in Ref. \cite{kozuma}. The authors of Ref. \cite
{kozuma} used two beams crossing each other at a small angle to produce a
one dimensional optical lattice with a very large period (about 5 times the
optical wavelength). Thus there was no Lamb-Dicke effect. The experiment,
however, showed clear oscillations which damped out much more slowly than
the inelastic scattering rate. Because the laser beam polarizations were
parallel to each other the atoms behaved nearly like two level systems in
the lattice and a Doppler cooling model was enough to show that this result
was not surprising. In the case of Doppler cooling one can show that the
mean occupation number of the harmonic oscillator $\left\langle
n\right\rangle $ is equal to the inelastic scattering rate divided by the
energy damping rate due to Doppler cooling. Thus $\langle n\rangle \gg 1$
implies a damping rate much smaller than the inelastic scattering rate, and
sidebands narrow compared to the inelastic scattering rate. This result was
also stated in Ref. \cite{cirac}.

For lattices in which the internal structure of the atom plays a significant
role in the cooling process (e.g. Sisyphus cooling), the Doppler model is
clearly not adequate. But as long as the motion of the atoms at the bottom
of the wells in an optical lattice is well described by a damped harmonic
oscillator, this damping rate determines the fundamental limit to the width
of the sidebands. (Note that Ref. \cite{raithel} observed a damping rate
which may be this fundamental rate. It is not yet clear, however, that the
motion can be described simply as a harmonic oscillator.) In what follows we
will show experiments in an optical lattice with Sisyphus cooling which
confirms these ideas. We show that the inelastic scattering rate, i.e. the
Lamb-Dicke parameter, has little to do with the width of the sidebands
observed.

\section{Intensity Correlation Spectroscopy}

As we have already mentioned, in correlation spectroscopy, one observes the
beating between the light emitted by different atoms in the source. Here we
will briefly review some of the ideas underlying intensity correlation
spectroscopy; the reader is referred to Refs. \cite
{jurczak1,jurczakth,westbrook2,cummins} for more detailed information. We
use a spectrum analyser to record the power spectrum $P_i(\omega )$ of a
detected photocurrent $i(t)$. The Wiener-Khinchine theorem states that $%
P_i(\omega )$ is proportional to the Fourier transform of the correlation
function of the photocurrent $\langle i(t)i(t+\tau )\rangle $. If we
consider the light as a classical field, the photocurrent is proportional to
the incident light intensity $I$ and our observation amounts to a
measurement of the normalized correlation function $g_2\left( \tau \right) $
of the scattered light intensity: 
\[
g_2(\tau )=\frac{\langle I(t)I(t+\tau )\rangle }{\langle I(t)^2\rangle } 
\]
where $\langle \rangle $ denotes a statistical average. Examples of the
spectral and time domain data are shown in Fig. 2. The spectrum (a) and the
correlation function (b) were taken using the same optical lattice, in one
case using a spectrum analyser and in the other using a digital correlator.
In 2(a) we have normalized the spectrum to that of a shot-noise limited
light source of the same average intensity, using a procedure described in
Ref.~\cite{jurczak1}. Thus a power density greater than 1.0 represents an
'excess noise power' which is due to the atoms. In 2(b) we plot $g_2\left(
\tau \right) -1$. For large values of $\tau $, this function tends to zero.
%
% Figure 2
%
\begin{figure}[tb]
\begin{center}
\includegraphics[width=\columnwidth]{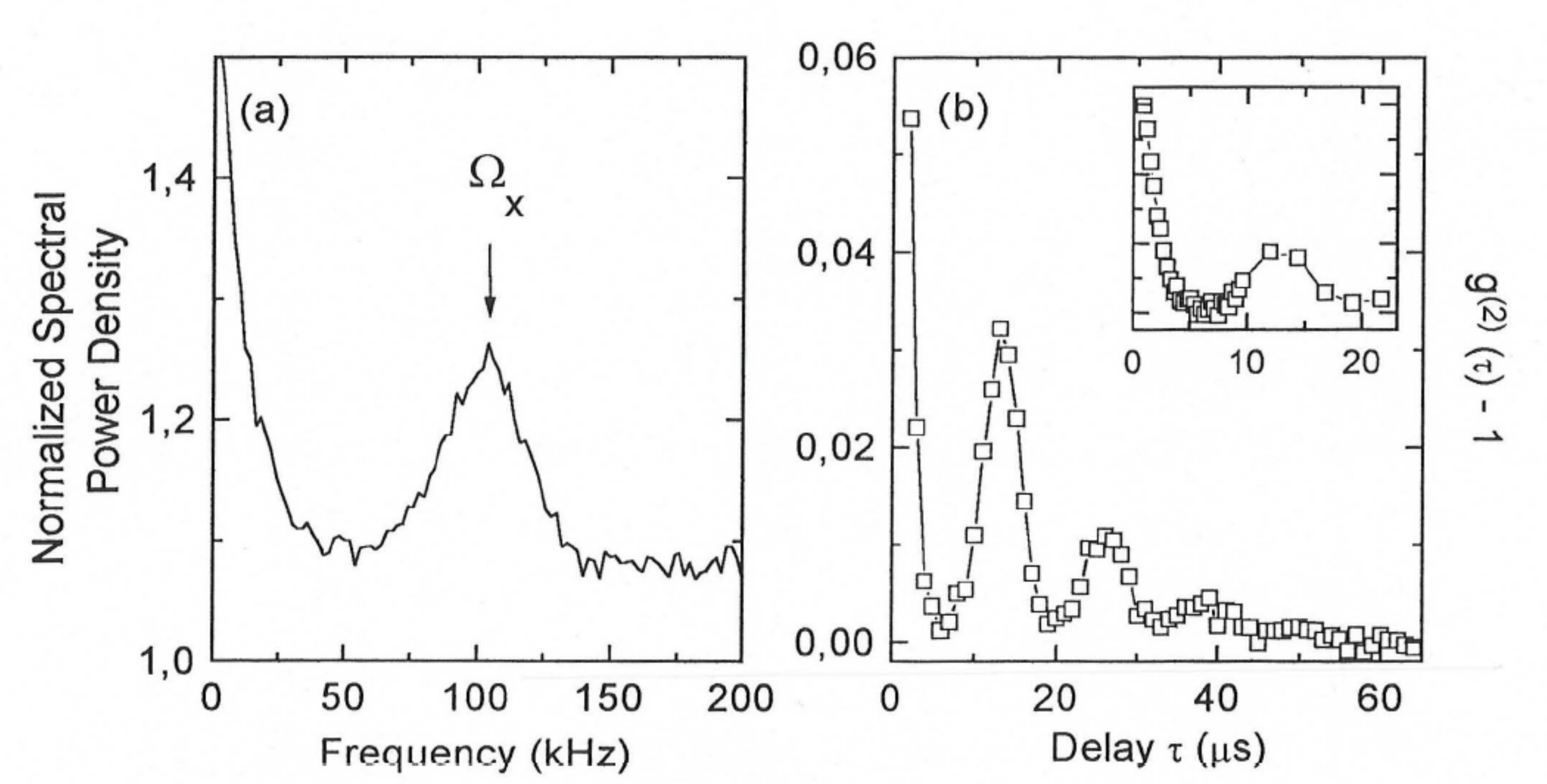}
\caption{
(a) Intensity correlation spectrum of atoms in an optical lattice,
showing a peak at zero frequency corresponding to elastic scattering and a
second peak (marked $\Omega _x$) at 100 kHz due to spontaneous Raman
scattering by atoms changing their level in the potential well. The angle of
the lattice was $\theta =30^{\circ }$, the observation direction was along $%
x $, the detetected light polarization was along $y$, and the detuning was $%
\Delta =-5\Gamma .$ (b) The autocorrelation function of the light as
recorded by a digital correlator under similar conditions.
}
\label{figure-2}
\end{center}
\end{figure}

If one is observing a large number of independent scatterers one can show
that the intensity correlation function is related to the correlation
function of the electric field $E$ by 
\begin{eqnarray}
g_{2}(\tau ) &=&1+\left| g_{1}(\tau )\right| ^{2}  \label{siegert} \\
g_{1}(\tau ) &=&\frac{\left\langle E^{-}(t)E^{+}(t+\tau )\right\rangle }{%
\left\langle E^{-}(t)E^{+}(t)\right\rangle }  \nonumber
\end{eqnarray}
where $E^{\pm }$ refers to the positive and negative frequency components of
the electric field. The Fourier transform of $g_{1}(\tau )$ is proportional
to the optical power spectrum, and this is the quantity that is measured in
heterodyne experiments. From Eq. \ref{siegert} it is easy to see why the
power spectrum of the photocurrent is proportional to the autoconvolution of
the optical power spectrum. The fact that we are dealing with an
auto-convolution, however, presents some difficulties in the quantitative
interpretation of the spectra. For example, the zero frequency peak in our
spectrum consists of the autoconvolution of the carrier plus the
autoconvolution of the sidebands. Thus its lineshape is complex. In what
follows we shall use a Montecarlo simulation to extract quantitative
information from our spectra.

\section{Description of the experiment}

We begin by collecting atoms in a vapor cell Magneto-Optical Trap (MOT) with
a Rb partial pressure of order 10$^{-8}$hPa. The lasers are tuned slightly
to the red of the F$_{g}$=3$\rightarrow $F$_{e}$=4 component of the D$_{2}$
resonance line of $^{85}$Rb ($\lambda =780$ nm, $\Gamma /2\pi =5.9$ MHz). We
can load of order 3$\times 10^{7}$ atoms into the trap with a time constant
of a few seconds. After loading the MOT, we switch off both the magnetic
field and the trapping beams and turn on the 4 lattice beams. These beams
come from a separate diode laser which is injected by the same master laser
as the trap laser. The density of atoms in the lattice is approximately 2$%
\times 10^{9}$ cm$^{-3}$. We allow the atoms to equilibrate in the lattice
for 5 ms before beginning the data acquisition period. All the lattice beams
enter the vacuum chamber by one of two large windows on the sides. Our
geometry permits angles $\theta $ as large as 40$^{\circ }$. This angle
could be changed fairly easily and in our experiments we compared values of
20$^{\circ }$, 30$^{\circ }$ and 40$^{\circ }$. These angles were measured
with a precision of approximately 1$^{\circ }$.

We measured the temperature of the atoms in our lattice by releasing the
atoms and monitoring their time of flight to a probe beam placed 18 mm below
the lattice (the $y$-axis is vertical). For a lattice angles of 30$^{\circ }$
we recorded the mean energy as a function of the depth of the potential. We
determined the potential depth by measuring the oscillation frequency of the
atoms and using Eq. \ref{oscfreq} and including a 10\% anharmonicity
correction (see below). The results are shown in Fig. 3. As expected, the
temperature depends linearly on the light shift. A linear fit to the data
gives: 
\begin{equation}
\frac{k_BT}2=0.17U_0+60E_R  \label{temp}
\end{equation}
This result corresponds closely to the results of Ref.~\cite{gatzke,kastberg}%
, which studied a 3D lattice at $\theta =45{{}^{\circ }}$ for Cs; both the
slope and the intercept seem to be somewhat higher in our case. This
discrepancy my be due to differences in the angle of the lattice. Although
Ref.~\cite{kastberg} reported that the temperatures were isotropic to within
20\%, our 2D simulations of the motion of atoms with a $J_g=1/2\rightarrow
J_e=3/2$ structure \cite{jurczakth,jurczak4} show, for lattice angles of 20${%
{}^{\circ }}$ or 30${{}^{\circ }}$, that the temperature is anisotropic with
the $x$ direction a factor of 1.5 hotter than the $z$ direction. The fact
that our measurements of the temperature along $x$ are above those reported
in Ref.~\cite{gatzke,kastberg} seems to bear this out \cite{bfield}.
%
% Figure 3
%
\begin{figure}[b]
\begin{center}
\includegraphics[width=12 cm]{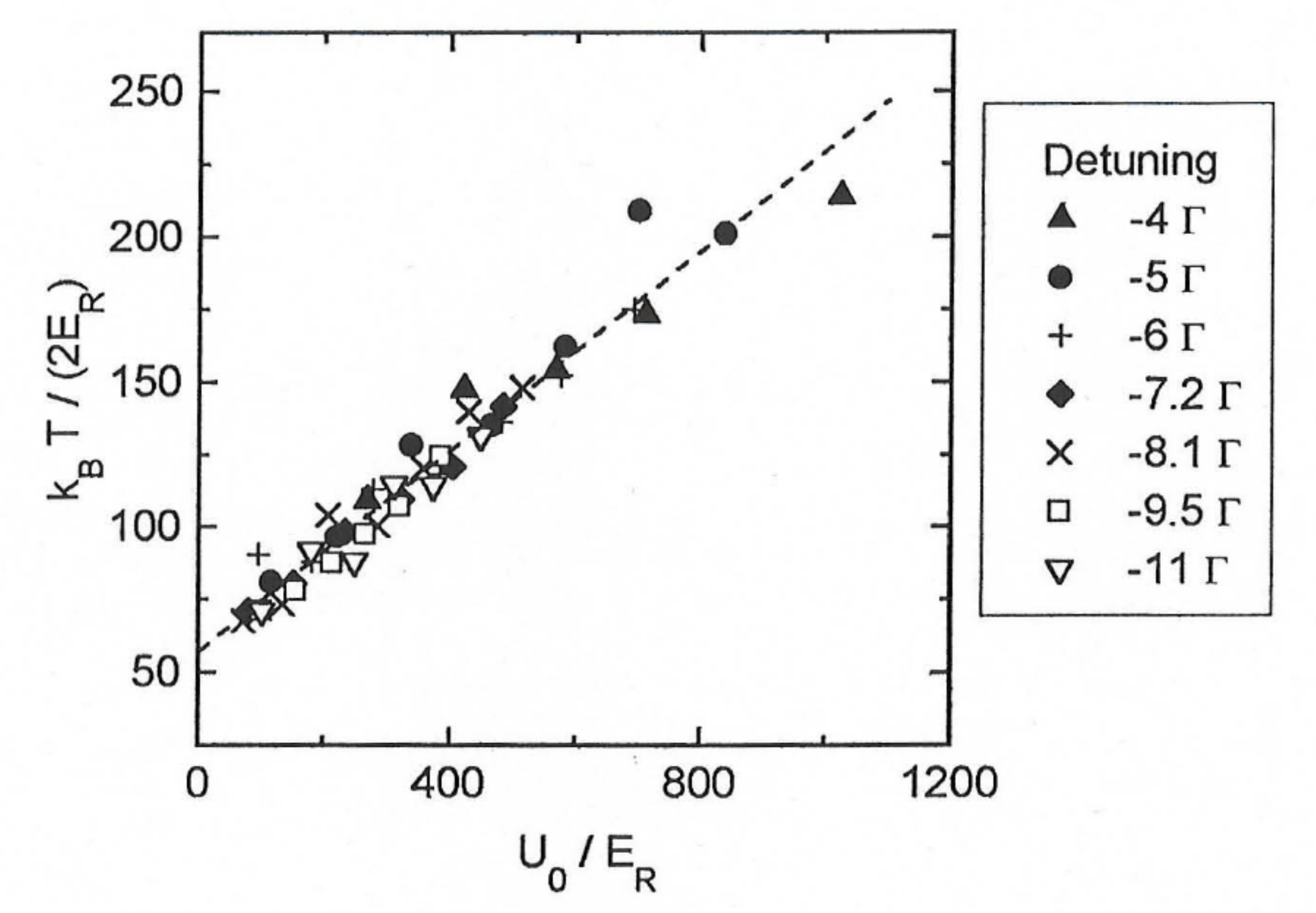}
\caption{
Temperature of atoms in a lattice with $\theta =30^{\circ }$
measured by time of flight as a function of the depth of the potential wells.
}
\label{figure-3}
\end{center}
\end{figure}
%
% Figure 4
%
\begin{figure}[t]
\begin{center}
\includegraphics[width=\columnwidth]{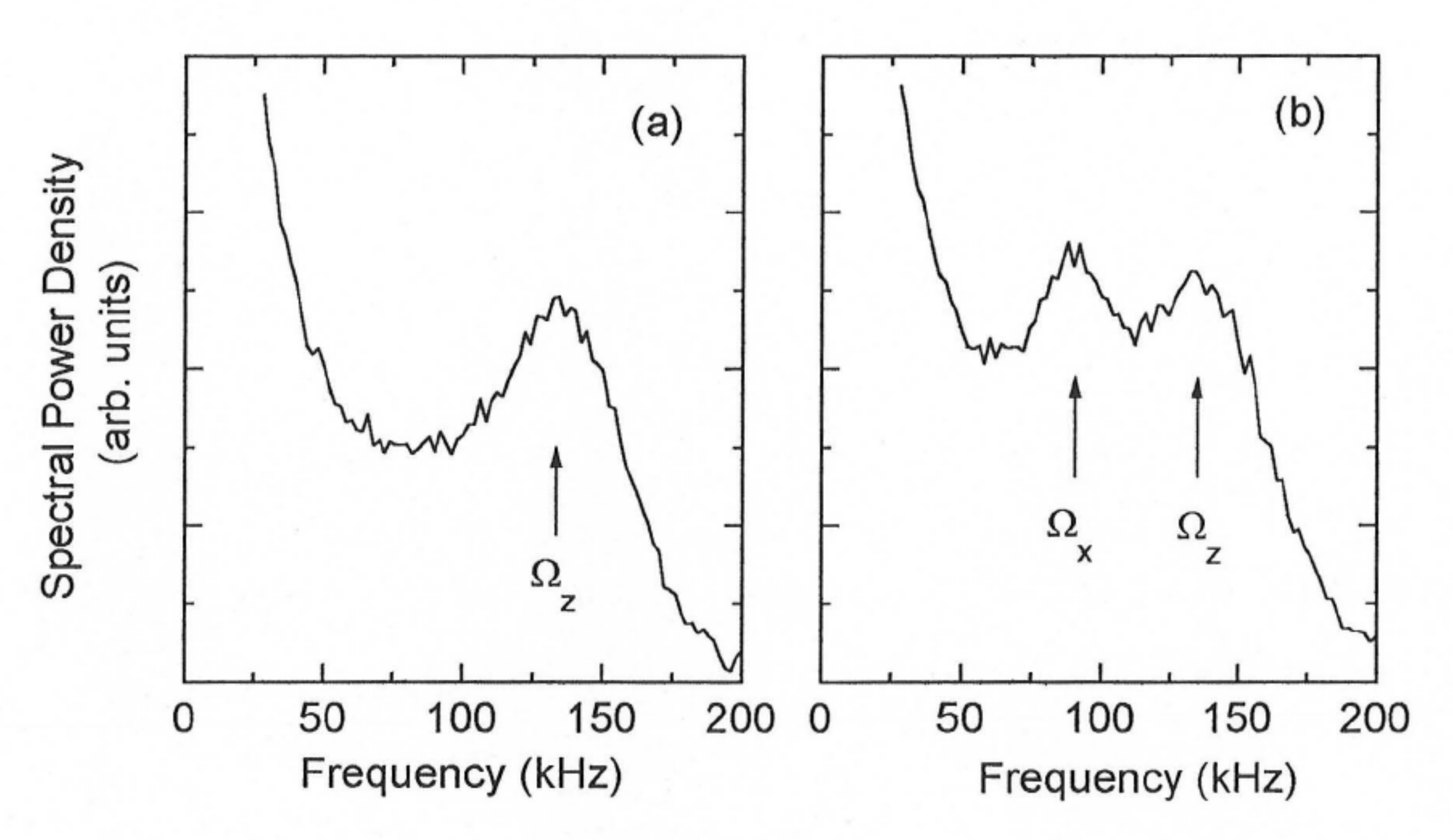}
\caption{
Intensity correlation spectrum of a lattice with $\theta
=40^{\circ }.$ (a) Observation along the $z$-axis, (b) Observation direction
in the $x-z$ plane at 10 degrees with respect to the $z$-axis. The detected
polarization was linear along $y$, and $\Delta =-5\Gamma $
}
\label{figure-4}
\end{center}
\end{figure}

In the correlation experiments, the light scattered by the atoms was
detected by avalanche photodiodes with integrated amplifiers and
discriminators. These signals were sent either to an FFT spectrum analyser,
or a digital correlator. The detection solid angle was approximately 10$%
^{-6} $ sr. Typical count rates were of order 10$^5$ to 10$^6$ s$^{-1}$. The
detectors were capable of a maximum count rate of 10$^6$ s$^{-1}$ without
significant saturation. The detector dark count rate was approximately 250 s$%
^{-1}$ and the background count rate, mostly due to scattering of the laser
beams by the vapor was of order 10$^4$ s$^{-1}$. When acquiring correlation
data, we typically loaded the MOT for 200 ms and used a data acquisition
period of 100 ms during which atom losses were negligible. The measurement
cycle was repeated and averaged 10$^2$ to 10$^5$ times.

\section{Results}

\subsection{Measurement of the oscillation frequencies}
%{\it Measurement of the oscillation frequencies. }
Figure 4 shows two
intensity correlation spectra taken from an optical lattice with $\theta =40{%
{}^{\circ }}$. In (a) the detection direction was the $z$ axis of Fig.1. In
(b) we used a detection angle of 10$^{\circ }$ from the $z$ direction. We
see that when one observes the light scattered along one of the axes of
symmetry of the lattice one sees only an oscillation in the direction of
that axis (see, however, Ref. \cite{jurczak3} for an exception). When
another observation direction is chosen one sees both of the
eigenfrequencies of the lattice. Note that the positions of the resonances
do not change when the observation angle is varied.
%
% Figure 5
%
\begin{figure}[t]
\begin{center}
\includegraphics[width=12 cm]{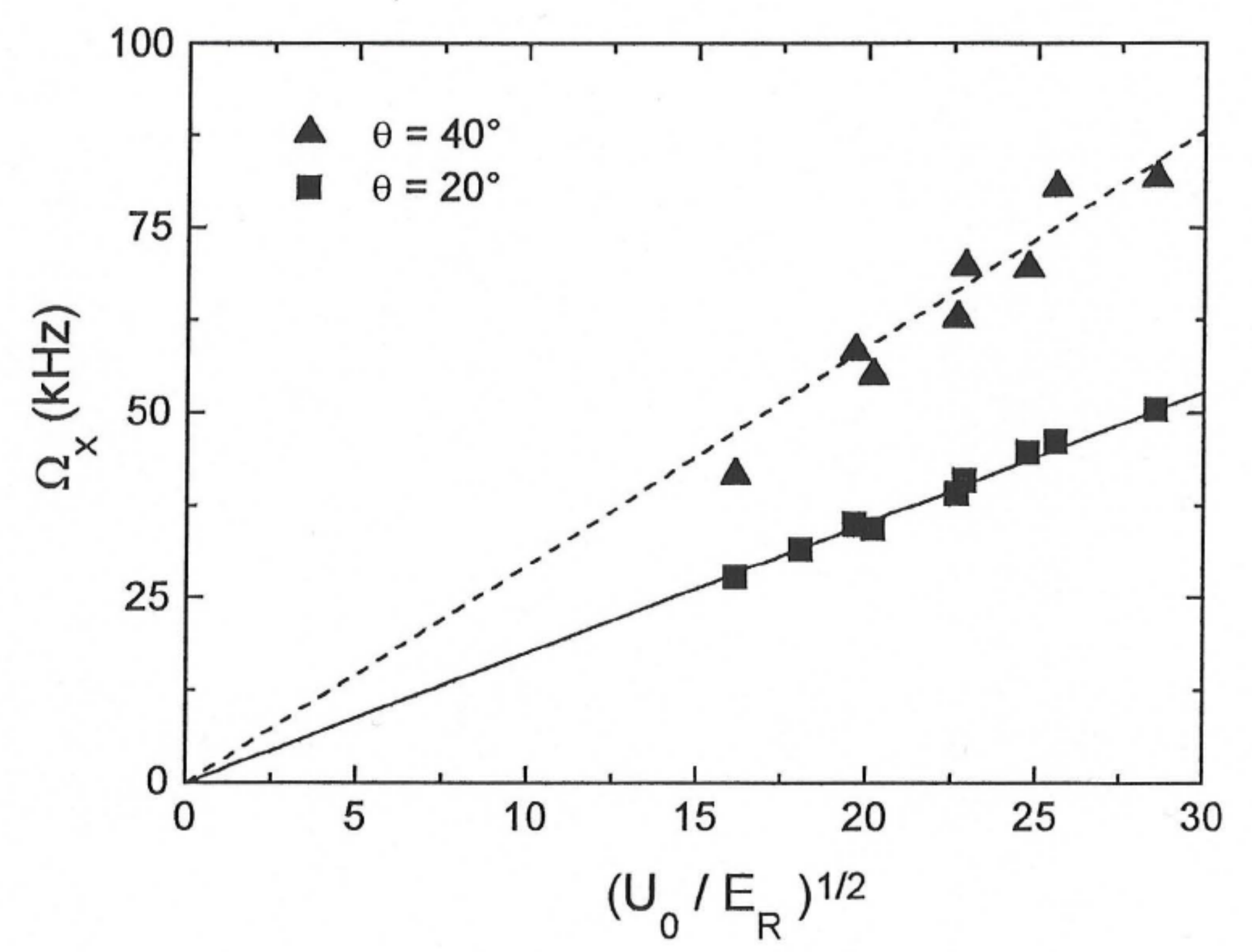}
\caption{
Measurements of the observed oscillation frequency as a function
of the square root of the potential well depth. We show the results for two
lattice angles. The dependence is linear as expected. The straight lines are
fits to the data (see text).
}
\label{figure-5}
\end{center}
\end{figure}

When the lattice angle is changed the two eigenfrequencies change according
to Eq. \ref{oscfreq}, smaller lattice angles leading to more widely spaced
frequencies. Fig. 5 shows our measurements of the dependence of oscillation
frequency predicted by Eq.\ref{oscfreq}. When we fit the data to a straight
line $\Omega _x=C\omega _R\sqrt{2E_R/U_0}\sin \theta $, we find $C=0.8$ and $%
C=0.9$ for $\theta =40^{\circ }$ and $\theta =20^{\circ }$ respectively. Our
precision is limited by our 10\% uncertainty in the laser intensity. Thus
our measurements are marginally consistent with Eq. 2 in which $C=1$.
However, we expect that the anharmonicity of the potential should reduce the
expected value of $C$ by of order 20\%\cite{gatzke}.

We have also made a systematic study of the relative positions of the two
resonances when they are measured simultaneously by observing the spectrum
at a small angle relative to $z$. The ratio should be much less sensitive to
uncertainties in the absolute measurement of $U_0$. A comparison between our
measurements and Eq.\ref{oscfreq} is shown in Table~\ref{frequencies} for the angles we
tested. As has already been demonstrated in \cite{verkerk2}, we find good
agreement between the predicted and measured ratios. A theoretical
investigation of the role of the potential anharmonicity, as well as the
possibly anisotropic temperature would be very useful to refine the
comparison with our results.
\begin{table}[t]
\caption{Measured and predicted ratios of the oscillation frequencies along $x$ and $z$
for different lattice angles.}
\begin{center}
\begin{tabular}{|c|c|c|}
\hline\
&\multicolumn{2}{c|}{$\Omega_x/\Omega_z$}\\
\cline{2-3}
Angle (degrees) & Predicted & Measured\\
\hline\hline
20&0.25&0.26\\ \hline
30&0.40&0.41\\ \hline
40&0.58&0.60\\ \hline
\end{tabular}
\end{center}
\label{frequencies}
\end{table}

\subsection{Width of the vibrational lines}
%{\it Width of the vibrational lines}. 
Our ability to vary the angles of the
lattice beams allows us not only to vary the oscillation frequency, but also
the degree of localization of the atoms in the wells. By localization we
mean the rms size of the atomic distribution in a well. Note that several
studies have shown \cite{marksteiner,gatzke,jurczakth} that for constant
geometry, the rms size of the atomic distribution is unchanged as the depth
of the potential is varied. This is because over a large range of parameters
the temperature of the atoms is proportional to the well depth. However, the
rms size varies with the size of the potential well, assuming that the mean
energy of the atoms remains roughly the same. Thus we can observe the
behavior of the width of the lines as the rms size is varied.

For the values of the lattice angle available to us, the largest variation
in the size of the potential is along $x$ or $y$. In Fig. 6 we show the
intensity correlation spectra along the $x$-direction for the angles $\theta 
$ of 20$^{\circ }$, 30$^{\circ }$ and 40$^{\circ }$ along with a contour
plot of the optical potential in the $x-z$ plane (for $y=0$). The contour
plots clearly show how the size of the potential varies as a function of
angle. The spectra were all taken for the same values of intensity and
detuning (i.e. of $U_0$ and the light scattering rate). The qualitative
features of these spectra can be resumed as follows: As the angle $\theta $
decreases, the oscillation frequencies become smaller, the intensities of
the sidebands relative to the peak at zero frequency become larger, the
width of the peaks gets slightly smaller, and more vibrational lines become
visible.
%
% Figure 6
%
\begin{figure}[t]
\begin{center}
\includegraphics[width=12 cm]{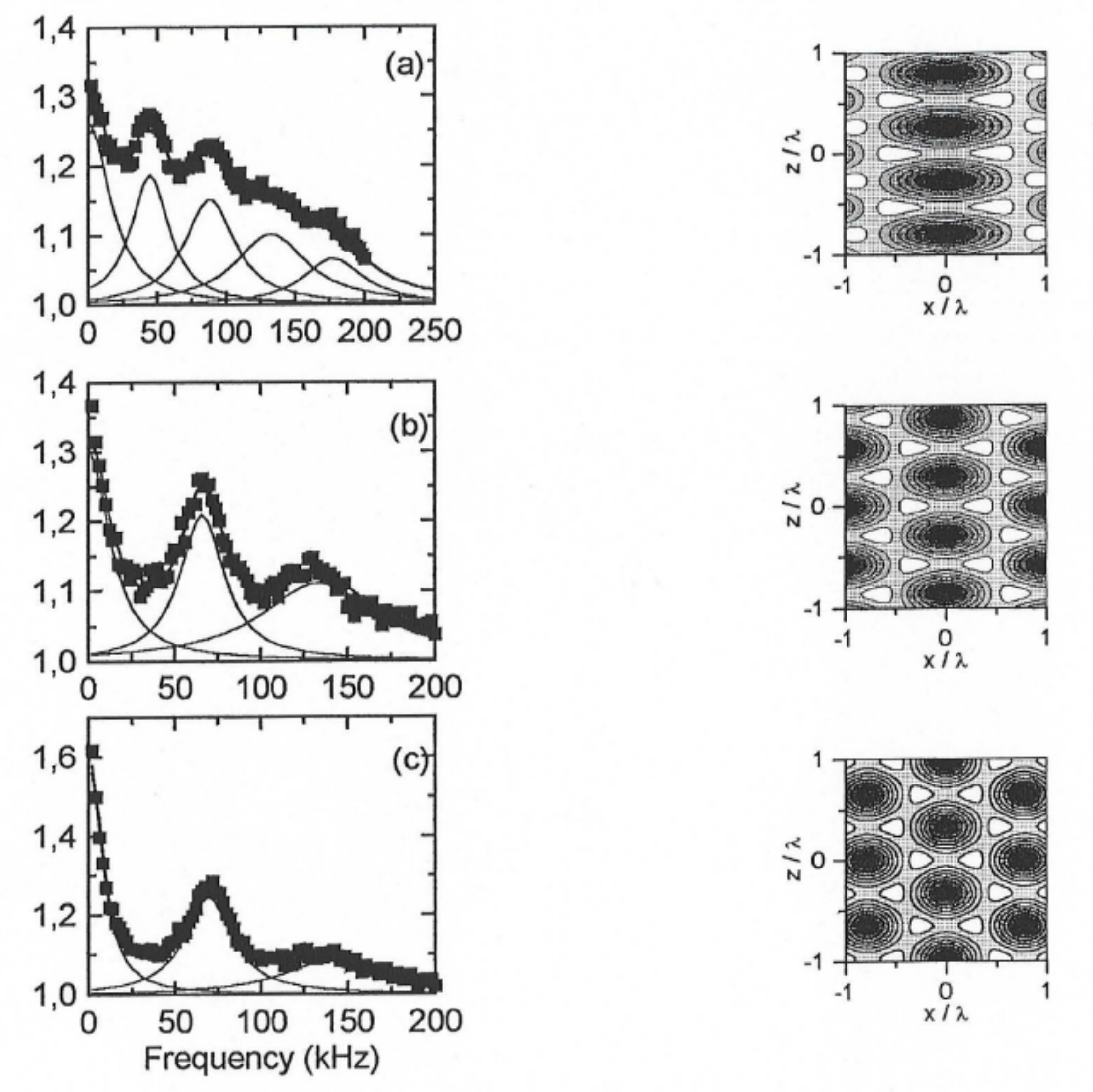}
\caption{
 Intensity correlation spectra observed for lattice angles of (a) 20%
$^{\circ }$, (b) 30$^{\circ }$ and (c) 40$^{\circ }$. For each spectrum $%
\Delta =-5\Gamma ,$ the observation direction was along $x$, and the
detected light polarization was along $y$. Next to each spectrum is shown a
contour map of the light shift potential, showing the anisotropy of the
potentials.
}
\label{figure-6}
\end{center}
\end{figure}

The qualitative interpretation of most these effects is also
straightforward. The vibrational frequencies are of course larger in the
smaller, stiffer potential wells. Secondly, as the stiffness of the well is
reduced, and if the average energy of the atoms remains constant, one also
expects that the rms size of the atom distribution increases thus reducing
the Lamb-Dicke effect. Therefore one expects the ratio of the intensities
radiated into the sidebands and to that in the elastic peak to increase. A
reduced Lamb-Dicke effect also favors the appearance of higher harmonics in
the vibrational spectra, although we note that if a third harmonic is
present in the data for 40$^{\circ }$ it would not appear in the observed
frequency range.

The last feature of the sidebands, their width, is difficult to understand
if one holds the view that the width of the sidebands is due to inelastic
scattering of the lattice light by the atoms. The fact that the ratio of the
height of the sidebands to that of the zero frequency peak increases as the
angle is decreased, shows that indeed the Lamb-Dicke effect is being
reduced, yet the width of the sidebands is going down. Note that Refs. \cite
{jurczak2,gatzke} have already pointed out that the widths of the sidebands
in optical lattices seem to be smaller that what would be predicted by the
inelastic scattering rate.

To be more quantitative, we would like a measurement of the value of $x_{rms}
$ the rms size of the distribution along the $x$ axis for the different
situations of Fig 6. One way to do this is to use the ratio ${\cal R}_{C}$
of the total power observed in the sidebands to that of the elastic peak in
the intensity correlation spectrum, to find the value of the Lamb-Dicke
parameter. Note that this ratio is not the same as the ratio ${\cal R}_{H}$
of the optical power in the sidebands to that in the elastic peak as
measured by the heterodyne method \cite{gatzke}. In the limit of strong
localization ($kx_{rms}\ll 1$) one can easily show that \cite{jurczakth}: 
\[
(kx_{rms})^{2}=\frac{{\cal R}_{C}}{2}={\cal R}_{H}.
\]
In our situation, however, $kx_{rms}$ is not small and therefore to relate $%
kx_{rms}$ to ${\cal R}_{C}$ we have simulated our experiment using our 2D
Montecarlo simulation of a $J_{g}=1/2\rightarrow J_{e}=3/2$ atom. We find
that even for angles of 30$^{\circ }$ and 40$^{\circ }$ the relation above
is approximately valid, and for $\theta =20^{\circ }$ $\frac{{\cal R}_{C}}{2}
$ gives an underestimate of $kx_{rms}$. The results of our simulations are
summarized in Table~\ref{comparison}. In both the simulation and the measurement we
estimate ${\cal R}_{C}$ from the spectrum by fitting the data to the sum of
several Lorentzians, one for each vibrational sideband as shown in Fig. 6.
Although this is not strictly the correct fitting function, (see \cite
{jurczakth}), the fits are quite good and we believe that the relative areas
give a good account of the relative areas of the elastic and inelastic peaks.

Another way to estimate $kx_{rms}$ is to take the measured value of the
temperature, Eq. 3, and use the formula for the potential energy, Eq. 2.
Then we use 
\begin{equation}
\frac{k_BT}2=\frac 12m(\Omega _xx_{rms})^2=\frac 12U_0(k\sin \theta
\;x_{rms})^2  \label{kTkxrms}
\end{equation}
where $m$ is the mass of the atom. For $\theta =30^{\circ }$ we can make
this comparison and as shown in the last column of Table~\ref{comparison} this estimate of $%
kx_{rms}$ is in reasonable agreement with the one from the spectrum.
\begin{table}[htdp]
\caption{A comparison of different methods of determing the Lamb-Dicke
parameter. The first two columns come from a numerical sumulation. They show
that for 30$^{\circ }$ and 40$^{\circ }$ the quantity ${\cal R}_{C}$, the
ratio of the inelastic to elastic peak intensities in the correlation
spectrum, can be used to determine the value of $(kx_{rms})^{2}$. The third
column shows the data deduced from Fig. 6, and the fourth column shows the
value of $(kx_{rms})^{2}$ determined by the temperature measurements shown
in Fig. 3 and using Eq. \ref{kTkxrms}. The internal consitency of these
results shows that for $\theta =30^{\circ }$ and $40^{\circ }$ the
Lamb-Dicke parameter increases as expected and that for $\theta =20^{\circ }$
our determination of $(kx_{rms})^{2}$ using ${\cal R}_{C}$ is likely to be
an underestimate.
}
\begin{center}
\begin{tabular}{|c|c|c|c|c|}
\hline\
&\multicolumn{2}{c|}{Simulation}
&\multicolumn{2}{c|}{Measurement}\\
\cline{2-5}
Angle (degrees)&$\mathcal{R}_\mathrm{C}/2$&$(kx_\mathrm{rms})^2$
&$\mathcal{R}_\mathrm{C}/2$&$(kx_\mathrm{rms})^2$\\
\hline\hline
20& 3.8 & 6.5 & 2.2 & ---\\ \hline
30& 2.5 & 2.5 & 1.8 & 2.2\\ \hline
40& 1.9 & 1.7 & 1.3 & ---\\ \hline
\end{tabular}
\end{center}
\label{comparison}
\end{table}

Thus we see that in all our experiments the Lamb-Dicke parameter is greater
than unity and thus one does not expect Lamb-Dicke narrowing to be present
at all. Nevertheless our light scattering rate is approximately $1.5\times
10^6$ s$^{-1}$ while the width of the sidebands is about $2\times 10^5$ s$%
^{-1}.$ In addition, while we increased the Lamb-Dicke parameter by about a
factor of 3, the width of the sidebands became slightly narrower. The fact
that the sideband width is so small even in the absence of Lamb-Dicke
narrowing supports the idea that the intrinsic or 'radiative' width of these
lines is governed by the damping rate of the atoms in the harmonic potential
and is quite small. We reiterate that the observed width is probably due to
the anharmonicity of the potential. Our Montecarlo simulations, in which the
atoms' motion is treated classically show comparable widths.

\subsection{Effects of polarization}
%{\it Effects of polarization. }
So far, we have not discussed the
polarization of the detected light. Note that in our laser beam
configuration, the electric field creating the lattice is $\sigma $
polarized (orthogonal to the $z$-axis) everywhere. When we observe the
spectrum along $z,$ the scattered light polarization is of course also $%
\sigma $, and we observe qualitatively the same spectrum regardless of
whether the detection polarization is $\sigma ^{+}$, $\sigma ^{-},$ $x$ or $%
y $ (see \cite{jurczakth} for more details). When observing the light
scattered along the $x$-axis, however, one can choose to observe either $%
\sigma $ or $\pi $ polarized light. In Figs. 2 and 6 we have shown data
taken with $\sigma $ polarization.
%
% Figure 7
%
\begin{figure}[t]
\begin{center}
\includegraphics[width=12 cm]{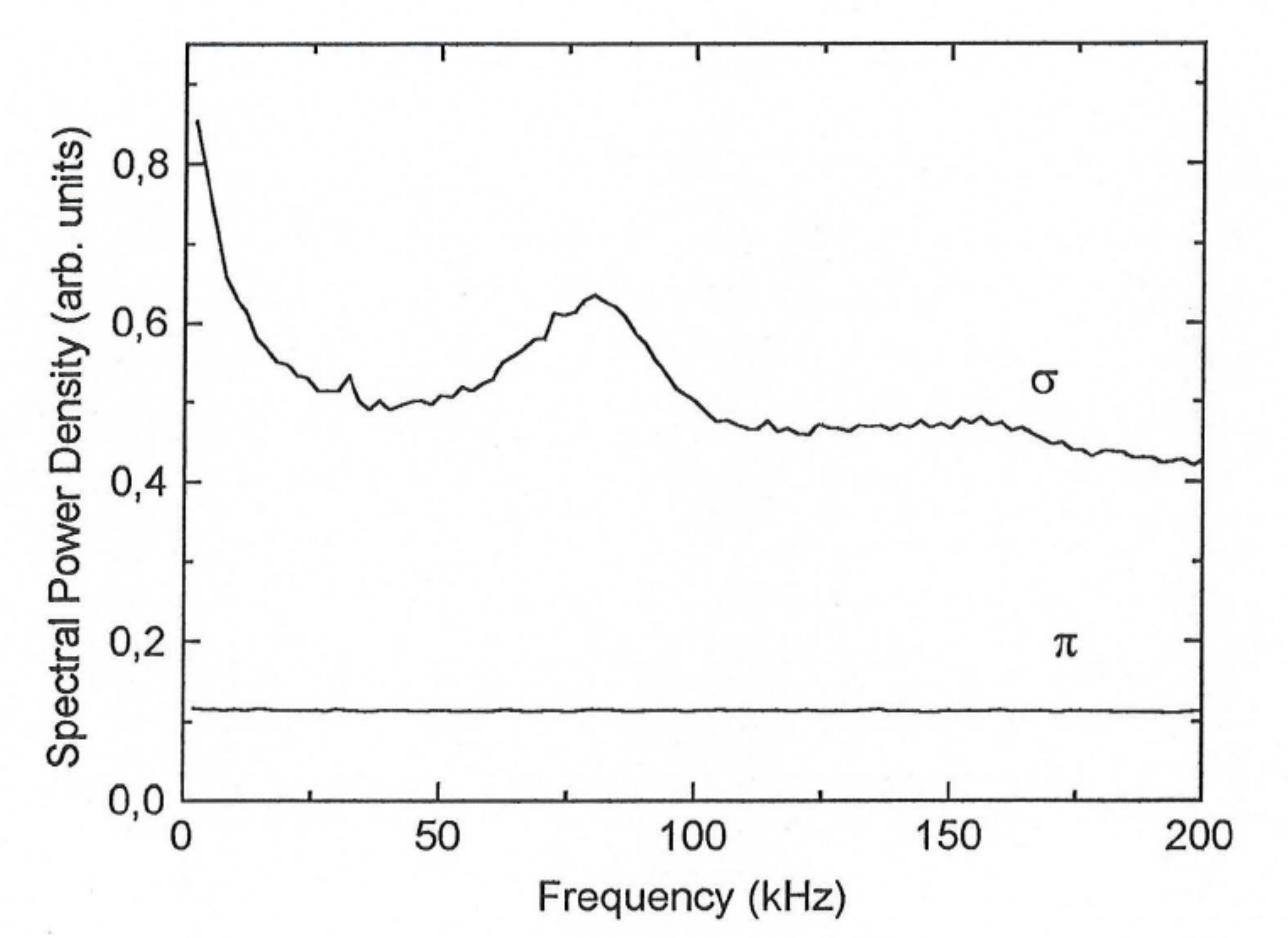}
\caption{
Spectral power density measured along the $x$-direction for two
polarization directions. The curve marked $\sigma $ corresponds to a
detection polarization along $y$. (as in Figures 2 and 6). The curve marked $%
\pi $ corresponds to detection of light polarized along the $z$ axis. For
both spectra $\theta =30^{\circ }$ and $\Delta =-5\Gamma .$
}
\label{figure-7}
\end{center}
\end{figure}

Surprisingly, the $\pi $ polarized spectrum shows {\it no} structure. There
is not even a zero frequency peak. We believe this is because by selecting $%
\pi $ polarized light, we have completely eliminated all elastic scattering:
absorption of a $\sigma $ polarized photon followed by the emission of a $%
\pi $ polarized one necessarily involves a change of the internal state of
the atom. Thus one expects this process to have an intrinsic width governed
by the inelastic light scattering rate \cite{polder}. There is also an
additional broadening mechanism, due to the fact that selecting the emitted
light polarization probably also corresponds to detecting light scattered by
atoms which are far from the minima of the potential wells. (In order to
emit a $\pi $ polarized photon, an atom in the $m_F=+3$ state must absorb a $%
\sigma ^{-}$ photon, and the $\sigma ^{-}$ intensity increases as one moves
away from the bottom of the $\sigma ^{+}$ well.) Thus we are selecting hot
atoms, many of which have an energy too large to be trapped by the potential
wells. This radiation will thus be Doppler broadended by the motion of these
atoms.

\section{Conclusions}

We have shown that intensity correlation measurements can give a great deal
of information about the motion of atoms in optical lattices. Our most
important results concern how the characteristics of the correlation
spectrum behave when the angles of the lattice are varied. The oscillation
frequencies obey a simple law which can be derived from the curvature of the
potential wells. Because the lattice angle governs the size of the potential
wells we also vary the Lamb Dicke parameter of the atoms in the lattice. We
clearly show that varying the Lamb-Dicke parameter has little impact on the
width of the motional sidebands. We hope our data will be useful for testing
future theoretical treatments, for example taking into account the 3D nature
of the problem. The question of what determines the structure of the
spectrum of the $\pi $ polarized light remains an interesting problem. To
get more experimental information it will be necessary to acquire data over
a much larger spectral bandwidth. Correlation measurements in the time
domain are well adapted to this type of measurement.

We thank J.-Y. Courtois for valuable discussions. W.D.P thanks NATO and the
CNRS for support and hospitality. G.B. was supported by E.U.grant ERB
CHBG-CT93-0436. This work was supported by DRET and by E.U.grant ERB
FMRX-CT96-0002.

$^1$present address: Laboratoire Kastler Brossel, 24 rue Lhomond, 75231
Paris Cedex 05, France.

$^2$present address: Institut f\"{u}r Quantenoptik, Univ. Hannover,
Welfengarten 1, 30167 Hannover, Germany.

$^3$permanent address: National Institute for Standards and Technology, PHY
A167, Gaithersburg MD, 20899 USA

\section{Appendix: Dynamics of localization using the time dependence of the
scattered light}

As we have discussed in Sec II, not only the depth of the light shift
potential, but also the local polarization of the electric field varies in
space. In particular, in our lattice the deepest points of the potential
correspond to $\sigma ^{+}$ or $\sigma ^{-}$ polarization, the quantization
axis being along $z$. This means that $\sigma $ polarized light is much more
likely to be emitted by localized atoms than $\pi $ polarized light. Thus if
one compares the intensities of the two light polarizations one should see a
significant difference. The information one gains in this way is akin to
what one gets in the experiments to observe Bragg reflection from a lattice 
\cite{birkl,weidemuller}. In our case, however, instead of appearing through
a Debye-Waller factor, localization of the atoms shows up simply as a signal
proportional to the average value of $\cos ^{2}(kz)$ of the atoms. In
practice it is difficult to use a steady state measurement of the scattered
light intensity to determine the mean position of the atoms, but, as with
recent Bragg diffraction experiments \cite{raithel,gorlitz}, we are able to
make a transient measurements to observe changes in the mean position of the
atoms when the conditions of the lattice are suddenly changed.

To illustrate this, consider the 1-dimensional lattice produced by 2
counterpropagating beams with orthogonal polarizations and an atom with a J$%
_{g}$=1/2 to J$_{e}$=3/2 transition \cite{courtois,dalibard}. We assume the
saturation parameter is small. The amount of $\sigma ^{+}$ or $\sigma ^{-}$
polarized light (i.e. polarized along a combination of $x$ and $y$) radiated
in all directions is proportional to \cite{jurczakth} 
\[
I_{\sigma \pm }\propto \left\langle \cos ^{2}(kz)\right\rangle _{m=\pm 1/2}+%
\frac{1}{9}\left\langle \cos ^{2}(kz)\right\rangle _{m=\mp 1/2}
\]
The $\langle \rangle $ brackets denote an average over all atoms in the
lattice with spins $m_{z}=$ $\pm $1/2. The second term in each bracket
represents the fraction of atoms in the $+(-)$ state which are excited by $%
\sigma ^{-}\left( \sigma ^{+}\right) $ light in the standing wave. The
numerical factors are due to the Clebsch Gordan coefficients and would be
different for other level structures. Similarly, for $\pi $ polarized light
(polarized along x), the total intensity in all directions is proportional
to 
\[
I_{\pi }\propto \frac{2}{9}\left( \left\langle \sin ^{2}(kz)\right\rangle
_{m=-1/2}+\left\langle \cos ^{2}(kz)\right\rangle _{m=+1/2}\right) 
\]

When the atomic distribution is homogeneous, the ratio of these two
quantities is 5. If on the other hand the atoms are sufficiently cold to be
highly localized then atoms with $m_{z}=$ $-$1/2 will tend to accumulate at
positions where $\cos ^{2}(kz)$ is large while atoms with $m_{z}=$ $+$1/2
will accumulate where $\sin ^{2}(kz)$ is large. This will increase $%
I_{\sigma ^{\pm }}$ and decrease $I_{\pi }$. In the limit of perfect
localization $I_{\pi }$ vanishes.

To apply this idea quantitatively to a 1D lattice with $^{85}$Rb atoms
(ground state F$_{g}$=3), and a detector which only observes the radiation
along say the $x$ direction, it would be necessary to include 2 additional
effects: the angular distribution of the radiated $\pi $ and $\sigma $
light, and the internal state distribution of the Rb atoms in the lattice.
In that case one could use the measured ratio $I_{\sigma ^{\pm }}/I_{\pi }$
to deduce the value of $\left\langle \cos {}^{2}(kz)\right\rangle $ . While
the first of these is straightforwardly calculated, the second of these
requires an independent measurement or calculation of the internal state
distribution of the Rb atoms. Finally in a 3D lattice it would also be
necessary to include the 3D dependence of the polarization of the electric
field. Thus it is quite difficult to use a simple intensity ratio to extract
an independent measure of the degree of localization of the atoms.
%
% Figure 8
%
\begin{figure}[t]
\begin{center}
\includegraphics[width=\columnwidth]{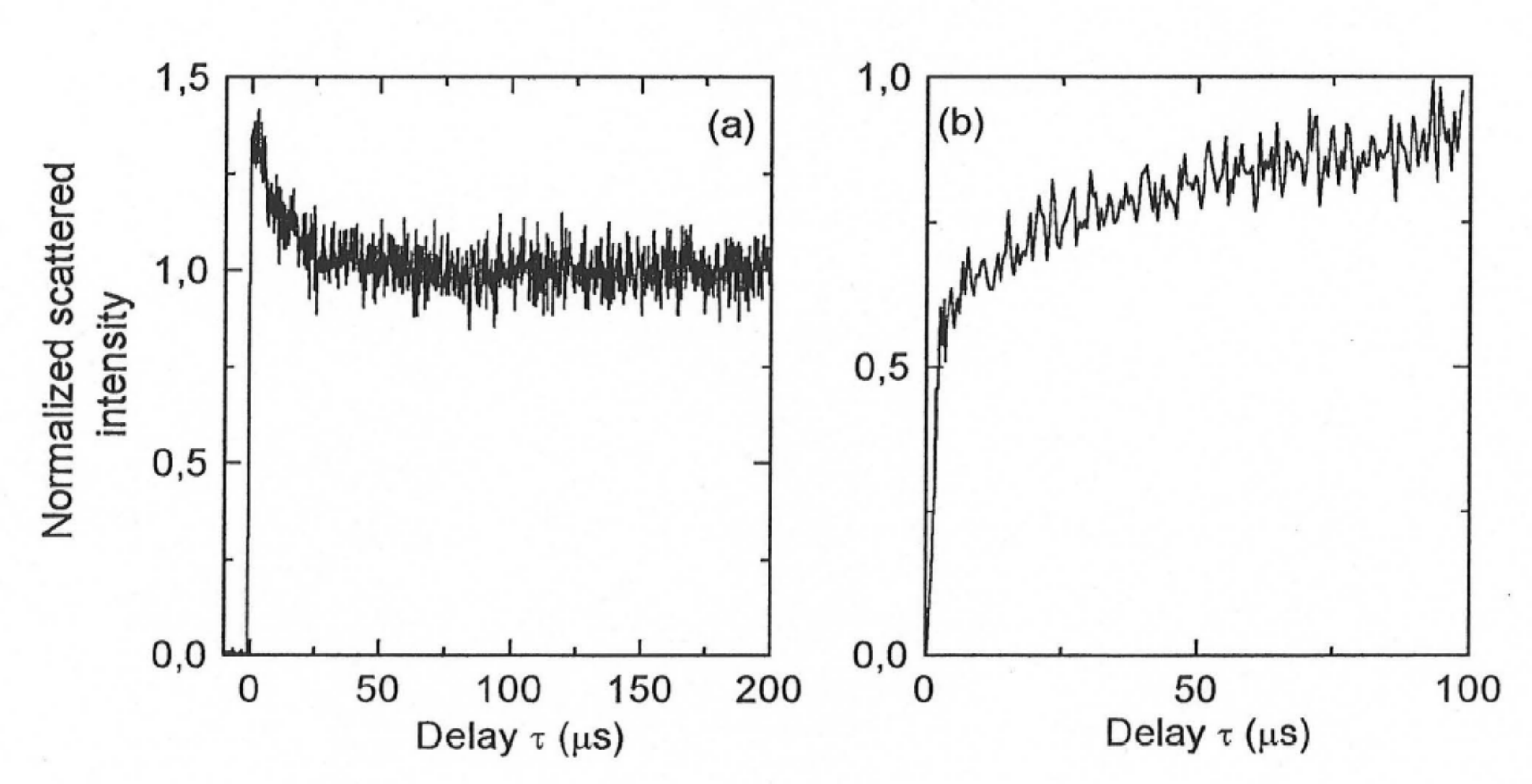}
\caption{
Intensity of (a) $\pi $ and (b) $\sigma ^{-}$ polarized light as a
function of time after switching on the optical lattice. The intensity is
normalized to its steady state value. The amount of light in the $\pi $
polarized component decreases with time as the atoms become increasingly
localized, while the $\sigma ^{-}$ intensity increases. For both figures $%
\theta =30^{\circ }$, $\Delta =-5\Gamma $. For curve (a) the observation
direction was $z$, and $U_0=760$ $E_R$. For $\left( \text{b}\right) $, the
observation direction was $x$ and $U_0=2100$ $E_R$.
}
\label{figure-8}
\end{center}
\end{figure}

In a transient experiment, however, one can use changes in the $\pi $ and $%
\sigma $ intensities to monitor changes in the degree of localization of the
atoms. We have undertaken such a measurement to demonstrate the sensitivity
of this technique. Our experiment proceeds as follows. After the MOT is
loaded we turn off the MOT beams and allow the atomic cloud to expand in the
dark for 2 ms. This time is long enough for the atoms to loose any trace of
their localization in the MOT beams. Next, we turn on the lattice beams and
monitor either the $\pi $ light intensity along the $x$ axis or the $\sigma
^{-}$ light along the $z$ direction. The intensities were monitored using an
avalanche photodiode and an multichannel scaler which plotted the number of
detected counts as a function of time after switching on the lattice beams.
The data are shown in Fig.8. Our measurement cycle lasted a few ms and was
repeated 10$^6$ times to achieve the signal to noise shown. As expected the $%
\pi $ intensity falls as the atoms become more localized, whereas the $%
\sigma ^{-}$ intensity increases with time.


\begin{references}
\bibitem{bfield}  Our high temperature may also be partly due to small
residual magnetic fields. Although we zeroed the field to better than 50 mG,
temperature data was not taken systematically as a function of the magnetic
field to ensure that we were operating at the minimum possible temperature.
\end{references}
\end{document}